\documentclass[aps,prl,twocolumn,groupedaddress]{revtex4-1}
\usepackage{amsmath}
\usepackage{graphicx}
\usepackage{color}

\begin{document}
\title{Spin polarisation by current}
\author{Sergey D. Ganichev$^1$, Maxim Trushin$^{1,2}$ and 
John Schliemann$^1$}
\affiliation{$^1$University of Regensburg, D-93040 Regensburg Germany,  $^2$University of Konstanz, D-78457 Konstanz, Germany}

\maketitle

\section{Introduction}

Spin generation and spin currents in semiconductor structures lie at the
heart of the emerging field of spintronics and are
 a major and still growing direction of solid-state research.
Among the plethora of concepts and ideas, current-induced spin polarization 
has attracted particular interest from both experimental and theoretical point of view, for 
reviews see 
Refs.~\cite{Silsbee04pR179,Rashba04p161201,Rashba05p137,Rashba06p0611194,Winkler06p0605390,Adagideli07p382,Dyakonov08Ivchenko,Kato08p031006,Ganichev08reviewmodphys08p1,Ganichev2014r,Sinova2015r,Nitta2015r,Jungwirth2016r,Schliemann2016r}.
In non-magnetic semiconductors or metals belonging to the  gyrotropic point groups\,\footnote{We remind that the gyrotropic point group symmetry makes no difference between certain components 
of polar vectors, like electric current or electron momentum, and axial vectors, like a spin or magnetic field, 
and is described by the gyration tensor~\cite{Dyakonov08Ivchenko,Landau,Nye}. 
Gyrotropic  media are characterized by the linear in light or electron wavevector 
$\mathbf k$ spatial dispersion resulting in  optical activity (gyrotropy) or 
Rashba/Dresselhaus band spin-splitting in 
semiconductor structures ~\cite{Dyakonov08Ivchenko,Nye,AgranovichGinzburg,Kizel,physchem,cardona_review}, 
respectively. Among 21 crystal classes lacking inversion symmetry, 18 are gyrotropic, from which 
11 classes are enantiomorphic (chiral) and do not possess a reflection plane or rotation-reflection 
axis~\cite{Dyakonov08Ivchenko,Kizel,physchem}. Three nongyrotropic noncentrosymmetric classes 
are T$_d$, C$_{3h}$ and $D_{3h}$. We note that it is often, but misleading, stated that 
gyrotropy (optical activity) can be obtained only in non-centrosymmetric crystals having 
no mirror reflection plane. In fact 7 non-enantiomorphic classes groups 
(C$_{s}$, C$_{2v}$, C$_{3v}$, S$_{4}$, D$_{2d}$, C$_{4v}$ and C$_{6v}$) are gyrotropic 
also allowing spin orientation by the electric current.}, see Refs.~\cite{Dyakonov08Ivchenko,Nye,AgranovichGinzburg,Kizel,physchem,cardona_review}, 
$dc$ electric current is generically accompanied by a non-zero
average nonequilibrium spatially homogeneous spin polarization and vice-versa. 
The latter phenomenon is referred to the spin-galvanic effect 
observed in GaAs QWs~\cite{Nature02} and other two-dimensional systems, see e.g. reviews~\cite{Winkler06p0605390,Dyakonov08Ivchenko,Ganichev2014r,Sinova2015r,Jungwirth2016r,Schliemann2016r,Ganichev03p935,GanichevPrettl,Ivchenko_book,Skinner}.
%
In low-dimensional semiconductor structures these effects are caused by 
asymmetric spin relaxation in systems with lifted 
spin degeneracy due to {$\mathbf k$}-linear terms in the Hamiltonian, where {$\mathbf k$} is the electron wave vector. 
In spite of the terminological resemblance, spin polarization 
by electric current fundamentally differs from the 
spin Hall effect~\cite{Rashba06p0611194,Winkler06p0605390,Adagideli07p382,Kato08p031006,Sinova2015r,Nitta2015r,Jungwirth2016r,Schliemann06p1015,Engel07,Hankiewicz09,Dyakonov71p467,Fabian08p565,Sinova08p45,Dyakonov08Dyakonov},
which refers to the generation of a pure spin current transverse to the charge 
current and causes spin accumulation at the sample edges.
The distinctive  features of the current-induced spin polarization
are, that this effect can be present in gyrotropic 
media only, it results in nonzero average spin polarization, and does not 
depend on the real-space coordinates. Thus, it can be measured in the whole 
sample under appropriate conditions. The spin Hall effect, in contrast, 
does not yield average spin polarization and  does not require gyrotropy, at least 
for the extrinsic spin Hall effect.

The ability of charge current to polarize spins in gyrotropic media has been predicted 
more than thirty years by Ivchenko and Pikus~\cite{Ivchenko78p640}. 
The effect has been considered theoretically for bulk tellurium crystals, where, almost to the same time, 
it has been demonstrated experimentally by Vorob'ev et al.~\cite{Vorobjev79p441}. In bulk tellurium
current-induced spin polarization is a consequence of the unique 
valence band structure of tellurium with hybridized spin-up and
spin-down bands (''camel back'' structure) and, in contrast to the spin polarization in 
quantum well structures, is not related to spin relaxation. In zinc-blende structure based  
QWs this microscopic mechanism of the current-induced spin polarization is absent~\cite{Aronov91p537,Vasko79}.
Vas'ko and Prima~\cite{Vasko79}, Aronov \& Lyanda-Geller~\cite{Aronov89p431}, and Edelstein~\cite{Edelstein89p233}
demonstrated that spin orientation by electric current is also 
possible in two-dimensional electron systems and is caused by asymmetric spin relaxation. 
%
Two microscopic mechanisms, namely scattering mechanism and precessional mechanism, based on Elliott-Yafet 
and D'yakonov-Perel' spin relaxation, respectively,  
were developed. The first direct experimental 
proofs of this effect were obtained  in 
semiconductor QWs
by Ganichev et al. in (113)-grown p-GaAs/AlGaAs QWs~\cite{Ganichev04p0403641,Ganichev06p127}, Silov et al.
in (001)-grown p-GaAs/AlGaAs heterojunctions~\cite{Silov2004p5929,Averkiev2005p1323} ,
Sih et al. in (110)-grown n-GaAs/AlGaAs QWs~\cite{Sih05p31} and Yang et al. in (001)-grown InGaAs/InAlAs QWs~\cite{Yang06p186605}, 
as well as in strained bulk (001)-oriented InGaAs and ZnSe epilayers by Kato et al.~\cite{Kato2004p176601,Kato05p022503} and Stern et al.~\cite{Stern2006p126603}, respectively.
The experiments include the range of optical methods such as Faraday rotation and linear-circular dichroism in
transmission of terahertz radiation, time resolved Kerr rotation
and polarized luminescence in near-infrared up to  visible spectral range. We emphasize 
that current-induced spin polarization was observed even at room 
temperature~\cite{Ganichev04p0403641,Ganichev06p127,Kato05p022503}. 
%
It has been demonstrated that depending on the point group symmetry 
electric current in 2D system 
may result in the in-plane spin orientation, like in the case of e.g. 
(001)-grown structures~\cite{Silov2004p5929,Averkiev2005p1323,Yang06p186605,Kato2004p176601,Kato05p022503,Stern2006p126603,Kuhlen,Stepanov,Norman,Hernandez2016}, 
or may additionally align spins normal to the 2DEG's plane. The latter take place in [llh]- or [lmh]-oriented structures, e.g. [113]-, [110]- and [013]-grown QWs (see~\cite{Ganichev2014r,Ganichev04p0403641,Ganichev06p127,Sih05p31,Golub2011}).
%
We also would like to note, that investigating spin injection from a ferromagnetic film into a two-dimensional electron gas, 
Hammar et al.~\cite{Hammar99p203,Hammar00p5024} used the  concept of a spin
orientation by current in a 2DEG (see also~\cite{Johnson98p9635,Silsbee01p155305}) to interpret their
results. Though a  larger degree of spin polarization was
extracted  the experiment's interpretation is complicated by other
effects~\cite{Monzon00p5022,Wees00p5023}. 
Later experiments on ferromagnet/(Ga,Mn)As bilayers~\cite{Skinner}, uniaxial (Ga,Mn)As
epilayers~\cite{Wunderlich},  metallic interfaces and surfaces~\cite{Fert,Zhang2014,Zhang2015,Sangiao,Nomura,Isasa} and interfaces between ferromagnetic films and topological insulators~\cite{Shiomi2014,Mellnik}   clearly demonstrated the ability 
of spin polarization by electric current in ferromagnetics- and metal-based structures.

The experimental observation of current-induced spin polarization 
has given rise to 
extended theoretical studies of this phenomenon in various systems
using various approaches and theoretical techniques.
These include the semiclassical 
Boltzmann 
equation~\cite{Golub2011,Vavilov05p195327,Tarasenko06p199,EngelRashba07p036602,Trushin07p155323,Raichev07p205340,Golub2013} 
%
derived from the
quantum mechanical Liouville equation~\cite{Culcer07p226601} and other
diffusion-type equations describing the dynamics of spin expectation 
values~\cite{Adagideli07p382,Burkov04p155308,Bleibaum06p035322,Korenev06p041308,Kleinert07p205326,Duckheim08p226602,Kleinert09p045317,Duckheim,Shen2014,Shen}. 
%
Other authors have performed important and fruitful studies
of the same issues using various Green's function techniques of
many-body physics~\cite{Rashba04p161201,Vaskobook05p662,Chaplik02p744,Inoue03p033104,Bao07p045313,Raimondi07p524,Liu08p125345,Gorini08p125327,Raimondi09}. 
The effect of external contacts to the
system and boundaries was studied explicitly in 
Refs.~\cite{Silsbee01p155305,Adagideli06p256601,Jiang06p075302}, 
and in Refs.~\cite{Jiang08p125309,Pletyukhov09p033303} the spin response of an electron gas to a microwave radiation was calculated. 
The influence of four terminal geometry has been studied in Ref.~\cite{Liu08p165316} using a numerical Landauer-Keldysh approach,
and weak localization corrections for current-induced spin polarization have been calculated in \cite{Guerci}.
The current-induced spin polarization has also  been
investigated in hole systems \cite{Liu08p125345,Zacharova06p125337,Garate2009}.
A search for efficient spin generation and manipulation
by all electrical means has given rise to theoretical analysis of current induced spin polarization in exotic regimes, like streaming caused by high electric fields~\cite{Golub2013,Golub2014} or very weak electron-impurity interaction \cite{Vignale2016},
in one-dimensional channels \cite{Averkiev,Trushin} and combined structures with  metal/insulator \cite{Tokatly}, ferromagnets/topological insulators~\cite{Garate2010,Yokoyama,Yokoyama2011,Luo2016}
ferromagnets/graphene~\cite{Manchon,Dyrda} or metal/semiconductor~\cite{Zhang2015,Borge2015}
interfaces.
%
%


Searching for publications on current-induced spin polarization one can be confused by the
fact that several different names are used to describe it. Besides current-induced spin polarization (CISP) this phenomenon is often referred to as the inverse spin-galvanic effect (ISGE), current-induced spin accumulation (CISA), 
the  magneto-electric effect (MEE) or kinetic magneto-electric effect (KMEE)
(this term was first used to describe the effect in non-magnetic conductors by Levitov et al.~\cite{Levitov85p133}), electric-field mediated in-plane spin accumulation. 
%
These variety of terms was extended further after the observation of the current-induced spin polarization at the interface between non-magnetic metals~\cite{Fert}. The authors introduce the new term - Edelstein effect (EE) - which, in followed up works, was modified to Rashba-Edelstein effect (REE) as well as to its inversion (IEE) corresponding to the spin-galvanic effect.  Despite the enormous diversity  of labels in all these cases we deal with one and the same microscopic effect: appearance of non-equilibrium spin polarization due to \textit{dc} electric current in the gyrotropic media with Rashba/Dresselhaus spin splitting of the bands. 
In our chapter we will use two of these terms: current-induced spin polarization and the inverse spin-galvanic effect.
%
%

%
\begin{figure}
\includegraphics[width=\columnwidth]{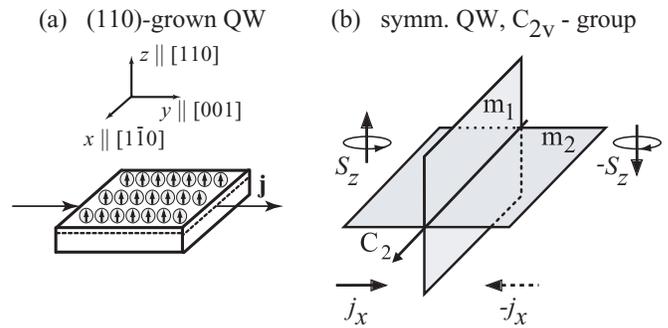}
\caption{(a) Electric current-induced spin polarization in symmetric 
(110)-grown zinc-blende structure based QWs.
(b) symmetry elements of  symmetrical QW grown in the $z ||$ [110] direction.
Arrows in the sketch (b) show reflection of the components of polar vector
$j_x$ and axial vector $S_z$ by the mirror reflection plane $m_1$. An additional 
reflection  by the mirror reflection plane $m_2$ does not modify the components 
of an in-plane polar vector $j_x$  as well as does not change the polarity of an out-plane axial
vector $S_z$. Thus, the linear coupling of $j_x$ and $S_z$ is allowed for
structures of this symmetry.}
\label{figure01}
\end{figure}

\section{Model}

Phenomenologically, the electron's averaged nonequilibrium spin ${\mathbf S}$ can be linked to the 
an electric current ${\mathbf j}$ by
\begin{eqnarray}
j_{\lambda} &=& \sum_{\mu} Q_{\lambda \mu} S_{\mu}\:, \label{Ch7currentequ22a}\\
S_{\alpha} &=& \sum_{\gamma}R_{\alpha \gamma} j_{\gamma} \,\, , \label{Ch7currentequ22b}
\end{eqnarray}
where {$\mathbf Q$} and {$\mathbf R$} are second rank pseudotensors.
Equation~(\ref{Ch7currentequ22a})
describes the spin-galvanic effect and Eq.~(\ref{Ch7currentequ22b})
represents 
the effect inverse 
to the spin-galvanic effect: an electron spin polarization induced by a 
$dc$ electric current.
We note the similarity of Eqs.~(\ref{Ch7currentequ22a}) and (\ref{Ch7currentequ22b}) characteristic for effects due to
gyrotropy: both equations linearly couple a polar vector with an  axial vector.
The phenomenological equation~(\ref{Ch7currentequ22b}) shows that the spin polarization can
only occur for those components of the in-plane components of $\mathbf j$ whose
transform as the averaged nonequilibrium spin ${\mathbf S}$ for all symmetry operations. 
Thus the relative orientation of the current direction and the average spin is
completely determined by the point group symmetry of the structure.
This can most clearly be illustrated by the example of a symmetric 
(110)-grown zinc-blende QW where an electric current along 
$x \parallel [1\bar{1}0]$ results in
a spin orientation along the growth direction $z$,  see Fig.~\ref{figure01}(a). 
These QWs belong to the point-group symmetry C$_{\rm 2v}$ and  contain, 
apart from the identity and a C$_2$-axis, a reflection plane
$m_1$ normal to the QW plane and $x$-axis and a refelction plane $m_2$ being parallel 
to the interface plane, see Fig.~\ref{figure01}(b).
The reflection in $m_1$ transforms the current component $j_x$ and the average spin 
component  $S_{z}$ in the same way 
($j_x \rightarrow -j_x$, $S_{y} \rightarrow -S_{y}$), see Fig.~\ref{figure01}(b).
Also the reflection in $m_2$ transforms these components in an equal way, 
and the sign of
them remains unchanged for this symmetry operation. Therefore, a
linear coupling of the in-plane current and the out-plane average spin is
allowed, demonstrating that a photocurrent $j_{x}$ can  induce the average spin polarization $S_z$.
In asymmetric (110)-grown QWs or (113)-grown QWs the symmetry is reduced to C$_{\rm s}$ and additionally
to the  $z$-direction, spins can be oriented in the plane of QWs\footnote{Note that 
for the lowest symmetry C$_{\rm 1}$, which contains no symmetry elements besides identity the 
relative direction between current and spin orientation becomes arbitrary.}.
Similar arguments demonstrate that 
in (001)-grown zinc-blende structure based QWs an electric current can results in an in-plane spin
orientation \textit{only}. In this case, the  direction of spins  depends  on the relative strengths of
the structure inversion asymmetry (SIA)~\cite{Bychkov84p78} and bulk inversion 
asymmetry (BIA)~\cite{Dyakonov86p110} resulting in an anisotropy of the current-induced spin polarization. 

\begin{figure}
\includegraphics[width=\columnwidth]{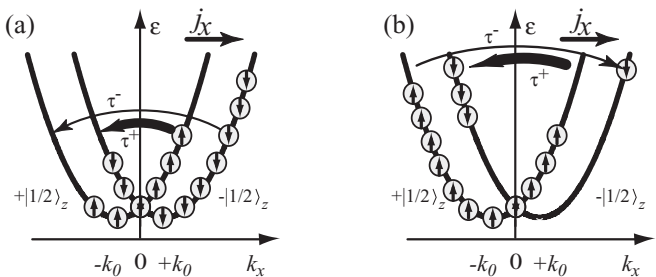}
\caption{ Current-induced spin polarization (a) and spin-galvanic effect (b) due to spin-flip scattering
in symmetric (110)-grown zinc-blende structure based QWs.
In this case only $\beta_{zx} \sigma_{z} k_x$ term are present in the Hamiltonian. The 
conduction subband is split into two parabolas with spin-up  $|+1/2\rangle_{z}$  and spin-down $|-1/2\rangle_{z}$ 
pointing in the $z$-direction. In (a) biasing along the $x$-direction causes an asymmetric in ${\mathbf k}$-space
occupation of both parabolas. In (b) nonequilibrium spin orientation along $z$-direction causes an electric current in $x$-direction.
After~\protect \cite{Ganichev04p0403641,Ganichev06p127}.}
\label{fig1isge}
\end{figure}

A microscopic model of the current-induced spin polarization~\cite{Ganichev04p0403641,Ganichev06p127} is sketched 
in Fig.~\ref{fig1isge}(a). To be specific  we consider an electron gas in symmetric (110)-grown zinc-blende 
QWs. The explanation of the effect measured in structures of other crystallographic orientation or 
in a hole gas can be given in a similar way. 
In the simplest case the electron's (or hole's) kinetic energy in
a quantum well depends quadratically on the in-plane wavevector components $k_x$
and $k_y$. In equilibrium, the spin degenerated  $k_x$ and $k_y$
states are symmetrically occupied up to the Fermi energy $E_F$. If
an external electric field is applied, the charge carriers drift
in the direction of the resulting force. The carriers are
accelerated by the electric field and gain kinetic energy until
they are scattered. A stationary state forms where the energy gain
and the relaxation are balanced resulting in a non-symmetric
distribution of carriers in ${\mathbf k}$-space yielding an electric current.  
The electrons acquire the average quasi-momentum
\begin{equation}
\label{momentum}
\hbar \Delta {\mathbf k}  = e  {\mathbf E} \tau_{ps}
\end{equation}
where ${\mathbf E}$ is the electric field strength, $\tau_{ps}$  is the
momentum relaxation time for a given spin split subband labeled by $s$, 
and $e$ is the
elementary charge. As long as spin-up and spin-down states are
degenerated in ${\mathbf k}$-space the energy bands remain equally
populated and  a current is not accompanied by spin orientation.
In QWs made of zinc-blende structure material like GaAs or strained
bulk semiconductors, however,
the spin degeneracy is lifted due to SIA and BIA~\cite{Bychkov84p78,Dyakonov86p110} 
and dispersion reads
\begin{equation}
E_{ks} = \frac{\hbar^2 \mathbf{k}^2}{2m^*} + \beta_{lm} \sigma_l k_m
\end{equation}
with the spin-orbit pseudotensor $\mathbf{\beta}$, the Pauli
spin matrices $\sigma_l$  and effective mass $m^*$ (note, the similarity to Eqs.~(\ref{Ch7currentequ22a}) and (\ref{Ch7currentequ22b})). 
The parabolic energy band splits into two subbands of opposite spin directions shifted in ${\mathbf
k}$-space symmetrically around ${\mathbf k} = 0$ with minima at $\pm k_0$. 
For symmetric (110)-grown zinc-blende structure based QWs  the spin-orbit interaction results in 
a Hamiltonian of the form  $\beta_{zx} \sigma_{z} k_x$,
the corresponding dispersion is sketched in Fig.~\ref{fig1isge}.
Here the  energy band splits into two subbands of $s_{z} = 1/2$ and $s_{z} = - 1/2$,
with minima symmetrically shifted in the ${\mathbf k}$-space along the
$k_x$ axis from the point $k = 0$ into the points $\pm k_0$, where
$k_0 = m^* \beta_{z'x}/ \hbar^2$.
As long as the spin relaxation is switched off the spin branches are equally 
populated and equally contribute to the current. Due to the band splitting, spin-flip
relaxation processes $\pm 1/2 \to \mp 1/2$, however, become different because
of the difference in quasi-momentum transfer from initial to final
states. In Fig.~\ref{fig1isge}(a) the $\mathbf k$-dependent
spin-flip scattering processes are indicated by
arrows of different lengths and thicknesses. As a consequence
different amounts of spin-up and spin-down carriers contribute to
the spin-flip transitions causing a stationary spin orientation.
In this picture we assume that the origin of the current-induced spin orientation is, as sketched in
Fig.~\ref{fig1isge}(a), exclusively due to scattering and hence 
dominated  by the Elliott-Yafet spin relaxation (scattering mechanism)~\cite{Averkiev02pR271}.
The other possible mechanism resulting in the current-induced spin orientation 
is based on the D'yakonov-Perel spin relaxation~\cite{Averkiev02pR271} (precessional mechanism). 
In this case the relaxation rate depends on the average 
$\Delta \mathbf k$-vector  equal to
${\bar k}_{1/2}= - k_0 + \langle \Delta k_x \rangle$
for the spin-up
and 
${\bar k}_{-1/2}= k_0 + \langle \Delta k_x \rangle$
for the
spin-down subband~\cite{Dyakonov86p110}. Hence, also for the D'yakonov-Perel
mechanism the spin relaxation becomes asymmetric in $\mathbf k$-space and 
a current through the electron gas causes spin orientation. 

Figure~\ref{fig1isge}(b) shows that not only phenomenological equations but also 
microscopic models of the current-induced spin polarization and  the spin-galvanic effect are similar.
Spin orientation in the $x$-direction causes the unbalanced population in spin-down
and spin-up subbands. As long as the carrier distribution in each
subband is symmetric around the subband minimum at $ k_{x_\pm}$ no
current flows. The current flow is caused by
{$\mathbf k$}-dependent spin-flip relaxation processes~\cite{Dyakonov08Ivchenko,Nature02,Ganichev03p935}. 
Spins oriented in the $z$-direction are scattered along $k_x$ from the
higher filled, e.g.,   spin subband $|+1/2 \rangle_z$, to the less
filled  spin subband $|-1/2 \rangle_z$. 
%
The spin-flip scattering rate depends
on the values of the wavevectors of the initial 
and the final states~\cite{Averkiev02pR271}.
Four quantitatively different spin-flip scattering events exist.
They preserve
the symmetric distribution of carriers in the subbands and, thus,
do not yield a  current.  
While  two of them have  the same rates, the other two,  sketched in
 Fig.~\ref{fig1isge} (b)  by bent arrows, are
inequivalent and generate an asymmetric carrier distribution
around the subband minima in both subbands. 
This asymmetric population results in a current flow along the
$x$-direction. Within this model of elastic scattering the current
is not spin polarized since the same number of spin-up and
spin-down electrons move in the same direction with the same
velocity. Like current-induced spin polarization also spin-galvanic effect
can result from the precessional mechanism~\cite{Dyakonov08Ivchenko} based 
on the asymmetry of the Dyakonov-Perel spin relaxation.

In order to foster the above phenomenological arguments and models, let us
discuss a microscopic description of such processes as introduced
in Ref.~\cite {Trushin07p155323}. This theory is based on an analytical
solution to the semiclassical Boltzmann equation, which does not include the spin relaxation time explicitly.
Instead one utilizes the so-called quasiparticle life time $\tau_0$,
which is defined via the scattering probability alone.
This is in contrast to the momentum relaxation time
$\tau_{ps}$, which, in addition, depends on the distribution function itself
and, therefore, should be selfconsistently deduced from the Boltzmann
equation written within the relaxation time approximation.
In the framework of a simplest model dealing
with the elastic delta-correlated scatterers 
the quasiparticle life time $\tau_0$ depends neither
on carrier momentum nor spin index, and in that way
it essentially simplifies our description.

Before we proceed let us first discuss the applicability
of the quasiclassical Boltzmann kinetic equation to
the description of the spin polarization in quantum wells observed in
\cite{Silov2004p5929,Averkiev2005p1323,Sih05p31,Yang06p186605,Kato2004p176601,Kato05p022503,Stern2006p126603}.
There are two restrictions which are inherited by
the Boltzmann equation due to its quasiclassical origin.
The first one is obvious: The particle's de Broglie length 
must be much smaller than the mean free path.
At low temperatures (as compared to the Fermi energy)
the characteristic de Broglie length $\lambda$ relates to the carrier concentration 
$n_e\sim \frac{m^* E_F}{\pi \hbar^2}$ approximately as
$\lambda\sim\sqrt{2\pi/n_e}$, whereas the mean free path $l$
can be estimated as $l\sim \frac{\hbar}{m^*}\sqrt{2\pi n_e}\tau_0$.
Thus, the first restriction can be written as
\begin{equation}
\label{first}
n_e \gg m^*/\hbar \tau_0.
\end{equation}
The second one is less obvious and somewhat more specific to our systems here, 
but still it is important:
The smearing of the spin-split subband due to
the disorder $\hbar/\tau_0$ must be much smaller than 
the spin splitting energy $E_{+k}-E_{-k}$.
The latter depends strongly on the quasimomentum, and
therefore, at the Fermi level it is defined by the carrier concentration.
As consequence, the concentration must fulfil the following
inequality
\begin{equation}
\label{second}
n_e \gg \hbar^2/8\pi\beta^2\tau_0^2\,,
\end{equation}
where $\beta$ is the spin-orbit coupling parameter.
This restriction can also be reformulated in terms
of the mean free path $l$ and spin precession length
$\lambda_s \sim \pi/k_0$.
Namely, $l$ must be much larger than $\lambda_s$ so that
an electron randomizes its spin orientation due to the spin-orbit precession
between two subsequent scattering events.
This restriction corresponds to the approximation
which neglects the off-diagonal elements of the 
non-equilibrium distribution function in the spin space.
Indeed, an electron spin can not only be in one of two possible
spin eigen states of the free Hamiltonian but in an arbitrary
superposition of them, the case described by the off-diagonal elements of
the density matrix. The latter is not possible in classical
physics where a given particle always has a definite position in the phase space.
Thus, we could not directly apply Boltzmann equation in its
conventional form for the description of the electron spin
in 2DEGs with spin-orbit interactions as long as the
inequality~(\ref{second}) is not fulfilled. 
Note, on the other hand, that all the quantum effects stemming
from the quantum nature of the electron spin
can be smeared out by a sufficient temperature
larger than the spin splitting energy $E_{+k}-E_{-k}$.
Thus, the room temperature $T_\mathrm{room}=25\,\mathrm{meV}$ 
being much larger than the spin-orbit splitting energy of the order of $3\,\mathrm{meV}$
(which is relevant for InAs quantum wells) makes the Boltzmann equation applicable for sure.
In order to verify whether the conditions (\ref{first}) and
(\ref{second}) are indeed satisfied, one can deduce the 
quasiparticle life time $\tau_0$ from
the mobility $\mu=e\tau_0/m^*$. Then, using spin-orbit coupling parameters 
in the range usually found in experiments~\cite{PRL2004ganichev,PRB2007giglberger,Meier08p035305}, the above inequalities turn out to be fulfilled.
In the case when inequality (\ref{second}) is not fulfilled, the theory based on spin-density
matrix formalism yields the result for CISP degree depending on the ratio
between energy and spin relaxation times \cite{Golub2011,Golub2012}. 
If energy relaxation is slower than Dyakonov-Perel spin relaxation, then
the electrically induced spin is given by Eq.~(\ref{tensor}) derived for
well-split spin subbands. This regime takes place at low 
temperatures. By contrast, spin density is twice larger in the opposite case of fast energy relaxation
which is realized at moderate and high temperatures.

In general, the Boltzmann equation describes the time
evolution of the particle distribution function $f(t,\mathbf{r},\mathbf{k})$,
in the coordinate $\mathbf{r}$ and momentum $\mathbf{k}$ space.
To describe the electron kinetics in presence of a
small homogeneous electric field $\mathbf{E}$ in a steady state
one usually follows the standard procedure
widely spread in the literature on solid state physics.
The distribution function is then represented as
a sum of an equilibrium $f_0(E_{s\mathbf{k}})$ and non-equilibrium $f_1(s,\mathbf{k})$ contributions.
The first one is just a Fermi-Dirac distribution,
and the second one is a time and coordinate independent non-equilibrium
correction linear in $\mathbf{E}$.
This latter contribution should be written down
as a solution of the kinetic equation, however,
it might be also deduced from qualitative arguments
in what follows.

Let us assume that the scattering of carriers is elastic, which means, 
above all, that spin flip is forbidden.
Then, the average momentum $\Delta {\mathbf k}$
which the carriers gain due to the electric field
can be estimated relying on the momentum relaxation time 
approximation from Eq.~(\ref{momentum}).
If the electric field is small (linear response)
then to get the non-equilibrium term $f_1(s,\mathbf{k})$ one has
to expand the Fermi-Dirac function $f_0(E_{s(\mathbf{k}-\Delta\mathbf{k})})$
into the power series for small $\Delta\mathbf{k}$ up to the
term linear in $\mathbf{E}$.
Recalling $\hbar \mathbf{v} = -\partial_{\Delta \mathbf{k}}E_{(s\mathbf{k}-\Delta\mathbf{k})}\mid_{\Delta \mathbf{k}=0}$
the non-equilibrium contribution $f_1(s,\mathbf{k})$ can be written as
\begin{equation}
\label{solution}
f_1(s,\mathbf{k})=-e\mathbf{E}\mathbf{v}\tau_{ps}
\left[-\frac{\partial f^0(E_{sk})}{\partial E_{sk}}\right].
\end{equation}
Since $f_1(s,\mathbf{k})$ is proportional to the derivative of the step-like
Fermi-Dirac distribution function, the non-equilibrium term substantially
contributes to the total distribution function close to the Fermi energy only,
and the sign of its contribution depends on the sign of the group velocity $\mathbf{v}$.
Note, that the group velocity $\mathbf{v}$ for a given energy and direction of  motion 
is the same for both spin split subbands and, therefore, non-equilibrium addition
to the distribution function would be the same for both branches as long as
$\tau_{ps}$ were independent of the spin index.
The latter is, however, not true, and $\tau_{ps}$ is different for two
spin split subbands, as depicted in Fig.~\ref{fig1isge}.
This is the microscopic reason why the non-equilibrium correction
to the distribution function gives rise to the spin polarization.

\begin{figure}[bt]
\includegraphics[width=0.9\columnwidth]{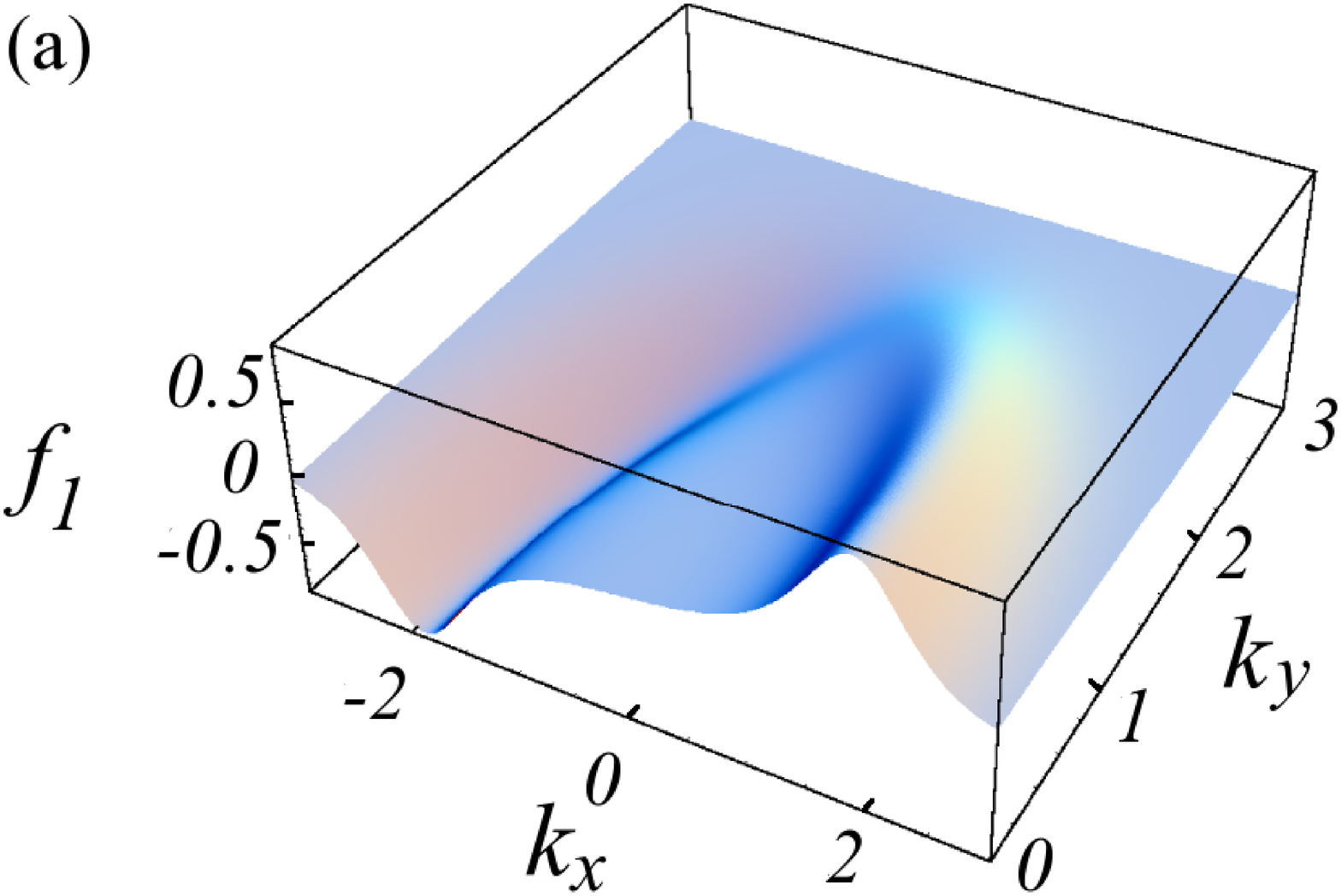}
\includegraphics[width=0.9\columnwidth]{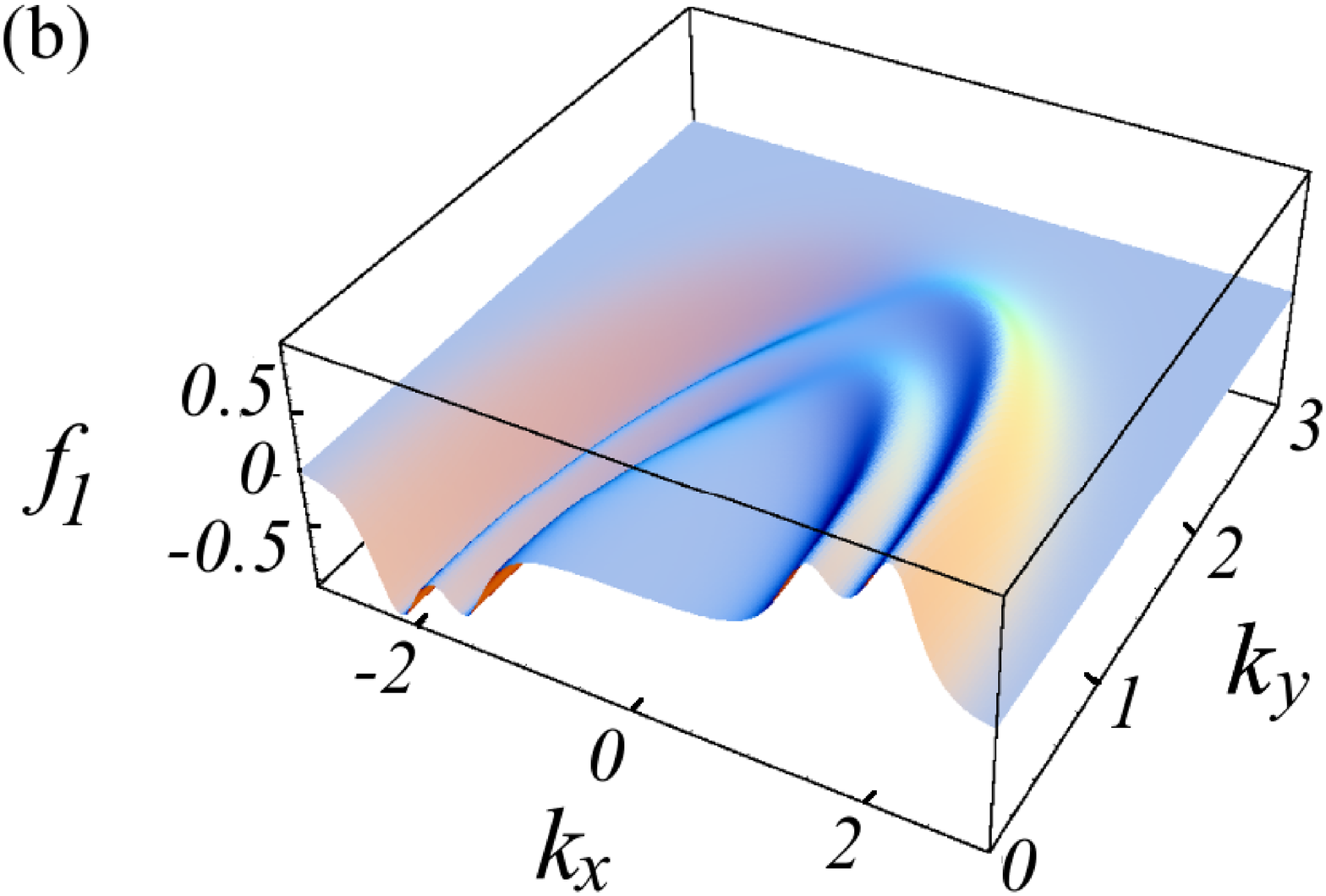}
\vspace*{8pt}
\caption{This plot shows schematically the nonequilibrium term of
the distribution function vs. quasimomentum for (a) spin-degenerate and (b)
spin-split subbands. The electric field is directed along the $x$-axis.
(a) Without spin splitting the non-equilibrium distribution function just provides
a majority for right moving electrons at the expense of the ones with opposite momentum.
(b) The spin-orbit splitting leads to the additional electron redistribution
between two spin split subbands. As one can see from the plot,
an amount of electrons belong to the outer spin split subband
is larger then for the inner one. Thus, the spins from inner and outer subbands are not
compensated with each other, and spin accumulation occurs.}
\label{fig3D}
\end{figure}

The non-equilibrium correction  can be also
found as an analytical solution of the Boltzmann equation 
\cite{Trushin07p155323},
and has the form
\begin{equation}
\label{elegant}
g_1(s,\mathbf{k})=-e\mathbf{E}\mathbf{k}\frac{\hbar\tau_0}{m^*} 
\left[-\frac{\partial f^0(E_{sk})}{\partial E_{sk}}\right].
\end{equation}
In contrast to Eq.~(\ref{solution}),  Eq.~(\ref{elegant}) contains 
momentum $\mathbf{k}$
and quasiparticle life time as a prefactor\footnote{Note that the well-known text book relation $\mathbf{v}=\hbar\mathbf{k}/m^*$
holds only in the absence of spin-orbit coupling.}.
Since the quasiparticle life time does not depend on the spin index,
and the quasiparticle momentum calculated for a given energy is obviously different for
two spin split subbands~Eq.~(\ref{elegant}), immediately allows us to read out
the very fact that the non-equilibrium contribution depends on the spin index,
see also Fig.~\ref{fig3D}.
Thus, the spins are not compensated with each other as long as the system is out of equilibrium.
Therewith one can see that the efficiency of the current-induced spin polarization
is governed by the splitting between two subbands, i. e. it is directly proportional
to the spin-orbit constants. One can also prove the later statements just  
calculating the net spin density  
\begin{equation}
\label{sd}
\langle S_{x,y,z} \rangle=\sum\limits_s\int\frac{d^2 k}{(2\pi)^2}
S_{x,y,z}(k,s) g_1(s,\mathbf{k}),
\end{equation}
with $S_{x,y,z}$ being the spin expectation values.

To conclude this section we would like to emphasize that
the two Eqs.~(\ref{solution}) and (\ref{elegant}) are just two ways
to describe the same physical content. One can think about
spin accumulation either in terms of the momentum relaxation
time $\tau_{ps}$ dependent on the spin split subband index,
or one  may rely on the exact solution Eq.~(\ref{elegant}) which
relates the spin accumulation to the quasimomentum difference
between electrons with the opposite spin orientations.

\section{Anisotropy of the inverse spin-galvanic effect}

Similar to the spin-galvanic effect~\cite{Dyakonov08Ivchenko,PRL2004ganichev,PRB2007giglberger} 
the current-induced spin polarization can be strongly anysotropic due to 
the interplay of the Dresselhaus and Rashba terms.
The relative strength of these terms is of general importance  because it is
directly linked to the manipulation of the spin of charge carriers
in semiconductors, one of the key problems in the field of spintronics.
Both Rashba and Dresselhaus couplings result in spin splitting of
the band  and give rise to a variety of spin-dependent phenomena
which allow us to evaluate the magnitude of the total spin
splitting of electron subbands~\cite{Ganichev03p935,GanichevPrettl,Fabian08p565,Averkiev02pR271,PRL2004ganichev,PRB2007giglberger,Meier08p035305,PRB1990luo,Knap1996p3912,Averkiev1999p15582,Tarasenko02p552,Tarasenko2,Schliemann03p146801,Winkler03,Zawadzki2003pR1,Zutic04review,Belkov08p176806,Lechner2009}.
Dresselhaus and Rashba terms can interfere in
such a way that macroscopic effects vanish though the indivi\-dual
terms are large~\cite{Averkiev02pR271,Tarasenko02p552,Tarasenko2,Schliemann03p146801}. For
example, both terms can cancel each other resulting in a vanishing
spin splitting in certain {$\mathbf k$}-space directions~\cite{Ganichev2014r,Schliemann2016r,Ganichev03p935,Schliemann03p146801}. 
This cancellation leads to the disappearance of an antilocalization~\cite{Knap1996p3912,Nitta}, 
circular photogalvanic effect~\cite{PRB2007giglberger}, magneto-gyrotropic effect~\cite{Belkov08p176806,Lechner2009},   
spin-galvanic effect~\cite{PRL2004ganichev} and current-induced spin polarization~\cite{Trushin07p155323},
the absence of spin relaxation in specific crystallographic
directions~\cite{Averkiev02pR271,Averkiev1999p15582},
the lack of Shubnikov--de Haas beating~\cite{Tarasenko02p552,Tarasenko2},
and has also given rise to a proposal for spin field-effect transistor
operating in the nonballistic regime~\cite{Schliemann03p146801}.

While the interplay of Dresselhaus and Rashba spin splitting may play a role 
in QWs of different crystallographic orientations 
here we focus on anisotropy of the inversed spin-galvanic effect in (001)-grown zinc-blende structure based  QWS.
For (001)-oriented QWs linear in wave vector part of Hamiltonian 
for the first subband reduces to
%
%
\begin{eqnarray} \label{biasia} 
{\cal{H}}^{(1)}_{\mathbf{k}} &=& \alpha(\sigma_{x_0} k_{y_0} - \sigma_{y_0} k_{x_0}) +
\beta(\sigma_{x_0} k_{x_0} - \sigma_{y_0} k_{y_0}) \:,
\end{eqnarray}
where the parameters $\alpha$ and $\beta$
result from the structure-inversion and bulk-inversion
asymmetries, respectively, and $x_0$, $y_0$ are the crystallographic
axes $[100]$ and $[010]$. 

To study the anisotropy of the spin accumulation one can just calculate
the net spin density from Eq.~(\ref{sd}), where
the integral over ${\mathbf k}$ can be taken easily
making the substitution $\varepsilon=E(s,k)$ and assuming that
$-\partial f^0(\varepsilon)/\partial \varepsilon=\delta(E_F-\varepsilon)$.
The rest integrals over the polar angle can be taken analytically, and
after some algebra we have
\begin{equation}
\label{tensor}
\langle \mathbf{S} \rangle = \frac{em^*\tau_0}{2\pi\hbar^3}
\left(\begin{array}{cc} \beta & \alpha \\  -\alpha & -\beta \end{array}\right)
\mathbf{E}.
\end{equation}
This relation between the spin accumulation and electrical current
can also be deduced phenomenologically applying Eq.~(\ref{Ch7currentequ22b}).

\begin{figure}
\includegraphics[width=0.9\columnwidth]{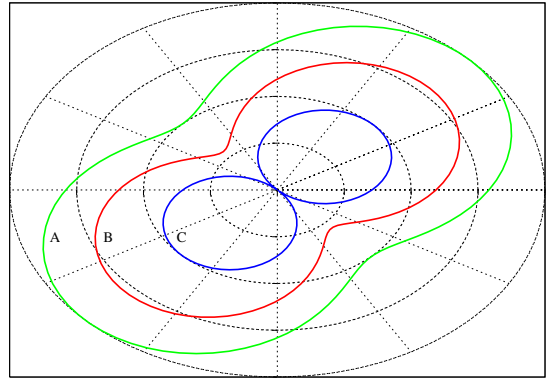}
\caption{\label{trushin} Anisotropy of spin accumulation (in arbitrary units) vs. direction
of the electric field in polar coordinates for different Rashba and Dresselhaus constants:
A --- $\alpha=3\beta$, B --- $\alpha=2.15\beta$, 
C --- $\alpha \sim \beta$ 
(after~\protect \cite{Trushin07p155323}).  
The curve B corresponds to the $n$-type InAs-based QWs investigated in~\protect \cite{PRL2004ganichev}. 
}
\end{figure}

The magnitude of the spin accumulation $\langle S \rangle=\sqrt{\langle S_{x}\rangle^2 + \langle S_{y}\rangle^2}$ 
depends on the relative strength of $\beta$ and $\alpha$ and varies after
\begin{equation}
\label{s_aver}
\langle S \rangle=\frac{eEm^*\tau_0}{2\pi \hbar^3}
\sqrt{\alpha^2+\beta^2+2\alpha\beta\sin(2\widehat{\mathbf{E}\mathbf{e}}_x)}.
\end{equation}
It is interesting to note, that $\langle S \rangle$ depends on the direction of the
electric field (see Fig.~\ref{trushin}), i. e. the spin accumulation is anisotropic.
This anisotropy reflects the relation between Rashba and Dresselhaus
spin--orbit constants. If either $\alpha=0$ or $\beta=0$ then
the anisotropy vanishes. 
%
In the opposite case of $\alpha\sim\beta$
the anisotropy reaches its maximum.  Note, that the case of $\alpha$ being
exactly equal to $\beta$ requires special consideration,
and the nonequilibrium distribution function turns out to be different from (\ref{elegant}).
Here, the effective magnetic field does not depend on the direction of motion,
and Dyakonov-Perel spin relaxation vanishes~\cite{Ganichev2014r,Schliemann2016r,Averkiev1999p15582,Schliemann03p146801}.
A similar effect has also been found in rolled-up heterostructures \cite{Trushin2007}, where Dyakonov-Perel spin relaxation also vanishes at a certain radius of curvature.
In such situations spin orientation by electric current becomes only possible due to other mechanisms of spin relaxation. Nevertheless, the anisotropy of spin relaxation
will reflect the anisotropy of the 
band structure and in our case of (001)-oriented QW will qualitatively correspond to the curve C in Fig.~\ref{trushin}.
%

The simple Drude-like relation between electrical current and field allows us
to write down a useful relation between the spin accumulation
and charge current density
\begin{equation}
\label{s-j}
\langle \mathbf{S} \rangle = \frac{m^{*2}}{2\pi\hbar^3 en_e}
\left(\begin{array}{cc} \beta & \alpha \\  -\alpha & -\beta \end{array}\right)
\mathbf{j}.
\end{equation}
%
Note that, in the coordinate system with the coordinate axes parallel to the mirror crystallographic planes
$x\parallel [1\bar{1}0]$ and $y\parallel [110]$, the Hamiltonian ${\cal{H}}^{(1)}_{\mathbf k}$
gets the form $\beta_{xy} \sigma_x k_y +
\beta_{yx}  \sigma_y k_x$ with $\beta_{xy} = \beta + \alpha$,
$\beta_{yx} = \beta - \alpha$, and
the relation between $\langle \mathbf{S} \rangle$ and 
$\mathbf{j}$ simplifies to
%
\begin{equation}
\label{s-j2}
\langle \mathbf{S} \rangle = \frac{m^{*2}}{2\pi\hbar^3 en_e}
\left(\begin{array}{cc} 0 & \beta+\alpha \\  \beta-\alpha & 0 \end{array}\right)
\mathbf{j}.
\end{equation}
Hence, for the spin accumulation $\langle S_{[110]} \rangle$ and $\langle S_{[1\bar{1}0]} \rangle$ 
provided by the electrical current along $[110]$ and $[1\bar{1}0]$ crystallographic directions, respectively
we obtain
\begin{equation}
\label{ab}
\frac{\langle S_{[1\bar{1}0]} \rangle }{\langle S_{[110]} \rangle  }=\frac{\alpha + \beta}{\beta-\alpha}.
\end{equation}
%
From equations (\ref{ab}) or (\ref{s-j2})   one can see that
the spin accumulation is strongly anisotropic if
the constants $\alpha$ and $\beta$ are close to each other.
As it was discussed earlier, the reason of such an anisotropy 
can be understood either from the spin-orbit splitting different for different direction
of the carrier motion, or from the anisotropic spin relaxation times 
\cite{Averkiev1999p15582}.

Equations (\ref{s-j2}) and (\ref{ab})   show
that measuring of the  spin polarization 
for current flowing in particular crystallographic directions 
one can  map the spin--orbit constants and deduce their magnitudes.
To find the absolute values of the spin--orbit constants
one needs to know the carrier effective mass $m^*$ and quasi-particle life time $\tau_0$.
The latter can be roughly estimated from the carrier mobility.
Indeed, though the band structure described by the Hamiltonian~(\ref{biasia}) is anisotropic,
the electrical conductivity remains isotropic that can be verified directly
computing the electrical current 
%
$$\mathbf{j}=-e\sum\limits_s \int \frac{d^2 k}{(2\pi)^2} \mathbf{v_s} g_1(s,\mathbf{k}).$$
%
Taking this integral in the same manner as in Eq.~(\ref{sd}) one can see that the anisotropy 
of the group velocity $\mathbf{v_s}$  is
compensated by the one stemming from the distribution function $g_1(s,\mathbf{k})$.
Therefore, the electrical conductivity is described by the Drude-like formula 
$\sigma=e^2 n_e \tau_0/m^*$ with 
$$n_e=\frac{m^* E_F}{\pi \hbar^2} + \left(\frac{m^*}{\hbar^2}\right)^2 \frac{\alpha^2+\beta^2}{\pi}$$
%
being the carrier concentration.
Using this Drude-like relation one can estimate
the quasiparticle life time as $\tau_0\sim m^* \mu/e$.
To give an example, an n-type InAs quantum well \cite{PRL2004ganichev} with
$\mu=2\cdot 10^{4}\, \mathrm{cm^{2}/(V\cdot s)}$ and
$m^*\sim 0.032m_0$ with $m_0$ being the bare electron mass yields 
the quasiparticle life time $\tau_0\sim 4\cdot 10^{-13}\, \mathrm{s}$.

Above we considered  the regime linear in the electric current.
Stronger electric fields provide a needle shape electron distribution known as ``streaming'' regime,
in which each free charge carrier accelerates quasiballistically in the ``passive'' region until reaching the optical-phonon energy, then emits an optical phonon and starts the next period of acceleration~\cite{Vosilyus67}. 
The inclusion of spin degree of
freedom into the streaming-regime kinetics gives rise to rich and interesting spin-related phenomena. In particular, the current-induced spin-orientation remarkably increases, reaching a high value $\simeq 2$\% in the electric field $\sim 1$ kV/cm. The spin polarization enhancement is caused by squeezing of the electron momentum distribution in the direction of drift \cite{Golub2013,Golub2014}. Note that further increase in the field spin polarization falls. 


Finally we would like to address another mechanism of the current-induced spin polarization and the spin-galvanic effect
proposed in Refs.~\cite{Tarasenko06p199,Tarasenko2008} and ~\cite{Golub2007}, respectively. It 
does not require the spin splitting of electron spectrum and is based on spin-dependent electron scattering by static defects or phonons which is asymmetric in the momentum space in gyrotropic structures~\cite{Averkiev02pR271,Ivchenko2004}. The spin polarization is then occurs due to asymmetric spin-dependent scattering followed by the processes of spin relaxation which can be
of both the Elliott-Yafet and the D’yakonov-Perel types. The scattering-related mechanism is expected to dominate at room temperature and/or electron gas of high density. It  may be responsible for the observed current-induced spin polarization in $n$-doped InGaAs epilayers~\cite{Norman}, where the anisotropy of both, current-induced spin polarization and spin splitting, was studied in the same structure  and the smallest spin polarization was surprisingly observed for the electric current applied in the crystal directions corresponding to larger spin splitting. While might be important in some particular cases, a 
detailed consideration of this mechanism is out of scope of our chapter.

\section{Concluding remarks}

We have given an overview of current-induced spin polarization in gyrotropic semiconductor
nanostructures. Such a spin polarization as response to a charge current,
%
may be classified as the inverse of the spin-galvanic effect, and  sometimes  
is called as magneto-electrical effect or Edelstein (Rashba-Edelstein) effect.
Apart from reviewing the experimental status of affairs, 
we have provided a detailed theoretical 
description of both effects in terms of a phenomenological model of
spin-dependent relaxation processes, and an alternative theoretical
approach based on the quasi-classical Boltzmann equation.
Two microscopic mechanisms of this effect, both involving \textbf{k}-linear band spin splitting, are known so far:
Scattering mechanism based on Elliott-Yafet spin relaxation and precessional mechanism
due to D'yakonov-Perel' spin relaxation. 
%
We note, that recently, another mechanism of the current-induced spin polarisation~\cite{Tarasenko2008}
and spin-galvanic effect~\cite{Golub2007}
which may dominate at room temperature has been proposed.
It is based on spin-dependent asymmetry of electron scattering and out
of the scope of the present chapter.
%
The relative direction between electric current and nonequilibrium spin for all mechanisms
is determined by the crystal symmetry. In particular, we have discussed the anisotropy 
of the inverse spin-galvanic effect in (001)-grown 
zinc-blende structure based QWs in the presence of spin-orbit interaction of both the Rashba and the
Dresselhaus type. Combined with the theoretical achievements derived from the Boltzmann
approach, the precise measurement of this anisotropy allows in principle to 
determine the absolute values of the Rashba and the Dresselhaus spin-orbit 
coupling parameter. 
Moreover, these interactions can be used for the manipulation of the magnitude and direction
of electron spins by changing the direction 
of current, thereby enabling a new degree of spin control.
We focused here on the fundamental physics underlying the spin-dependent transport of carriers in semiconductors, 
which persists up to room temperature~\cite{Ganichev04p0403641,Ganichev06p127,Kato05p022503}, 
and, therefore, may become useful in future semiconductor spintronics.


\section{Acknowledgments}

We thank E.L.~Ivchenko, V.V.~Bel'kov, D.~Weiss, L.E.~Golub,  and S.A.~Tarasenko for helpful discussions. This work is supported by 
the DFG (SFB 689) and the Elite Network of Bavaria (K-NW-2013-247).

\bibliography{spin-upd}

\end{document}